\documentclass{amsbook}

\newcommand{\chpsummary}[1]{#1}

\usepackage{graphicx,natbib,amsmath}

\newcommand{\ext}{jpg}
\usepackage{sidecap}

\begin{document}
\chapter{Extragalactic Radio Sources as a Piece of the Cosmological Jigsaw}

\label{mghk}

\chaptermark{Extragalactic Radio Sources}
\vspace{0.1in}




\vspace{0.1in}


\chpsummary{Early on,
extragalactic radio sources have 
pointed to a cosmologically evolving Universe. They were also an
important piece of evidence for
the existence of supermassive black holes, now thought to be a key component of galaxies. The 
observation that the power of radio sources increases with redshift, whereas the cosmological
assembly of mass proceeds vice versa means that radio sources have their strongest impact 
in the early Universe. Our simulations suggest that radio sources heat hot halo gas, boost star formation
in disc galaxies and other cold gas in the vicinity, possibly filaments, by a “surround and squash” 
mechanism. They might cause gaseous outflows in connection with stellar feedback. 
This might be an important mode of star formation for forming massive galaxies. Analysis of the
jet-environment interaction may provide insights into black-hole physics and jet formation, e.g.,
rotational energy extraction (Blandford-Znajek) or how frequent black-hole binaries
or multiple systems are. The former relates to fundamental questions about the nature of 
black holes. The latter is expected from hierarchical cosmology.
Extragalactic radio sources thus continue to corroborate the cosmological picture and lead the 
way towards new, exciting discoveries.}



\section{Introduction} \label{mghk-s:intro}
\begin{figure}
  \centering
  \includegraphics[width=0.9\textwidth]{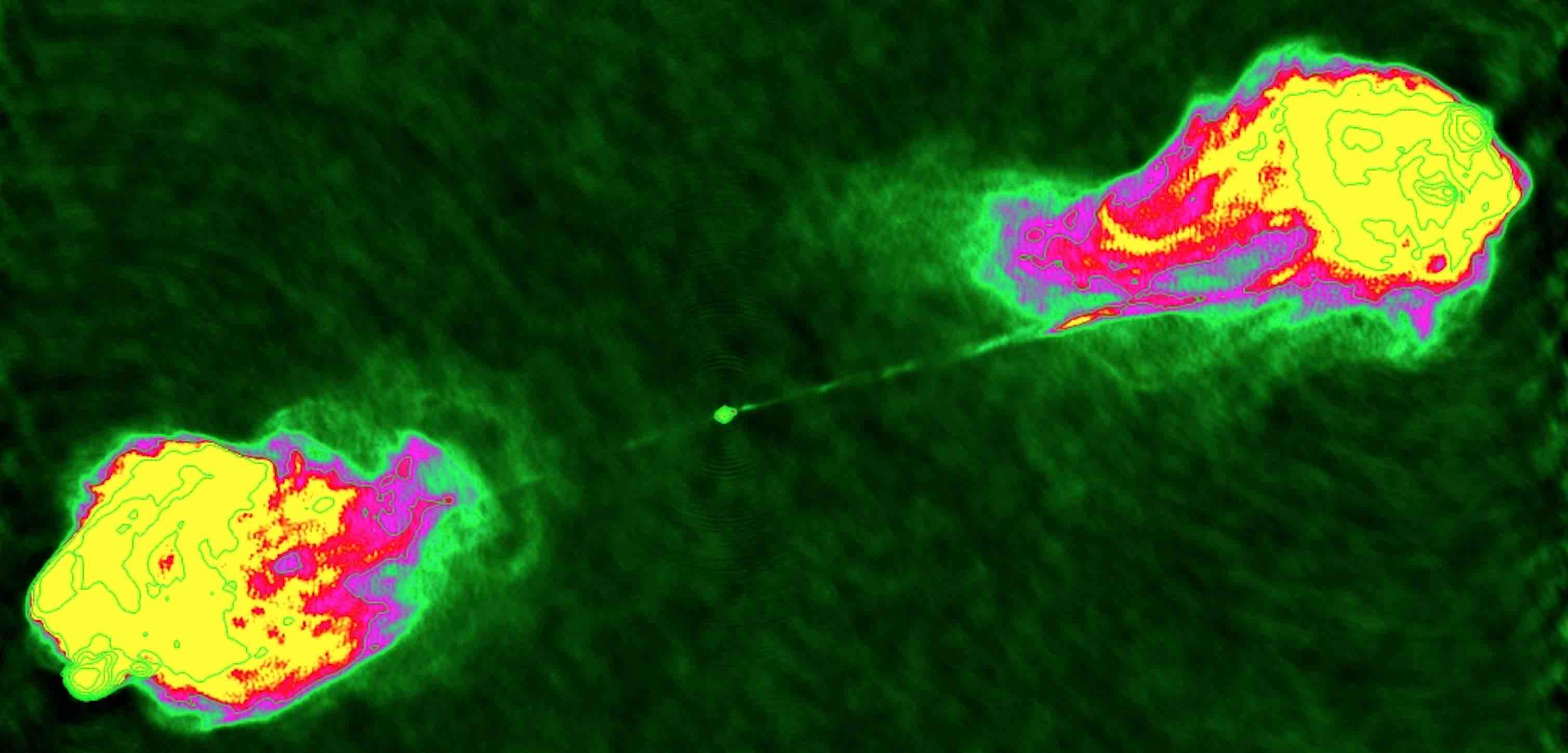}
  \caption{\small Very Large Array 5 GHz image of the first identified extragalactic radio source, Cygnus~A \citep{mghk-CB96}. 
  	The colour scale was adjusted to emphasise the faint jets. Contours reveal the bright hotspots.
  	The radio map was kindly provided by Chris Carilli and Rick Perley.}
   \label{mghk-f01}%
\end{figure}

Extragalactic radio sources were first identified in the 1950s and very soon opened up a new perspective on the Universe
\citep[for a review]{mghk-norris17}.
They were much more powerful than galaxies and the fact that they could be resolved with radio interferometers meant that they were
bigger than galaxies, perhaps related to galaxy mergers \citep{mghk-bm54}.
The first identified extragalactic radio source, Cygnus~A (Fig.~\ref{mghk-f01}), was optically identified with a central member of a galaxy cluster
\citep{mghk-bm54}. This proved to be no accident: Observationally, central cluster galaxies are frequently associated
with powerful radio sources, especially if they have a dense and rapidly cooling intra-cluster medium \citep{burns90}.
The dense gas is crucial to confine the radio emitting plasma, such that the observed high energy densities may be produced
\citep[e.g.][]{mghk-KAD97,mghk-HK13}. Powerful radio sources are hence observed in environments
of enhanced galaxy density \citep{mghk-ls79}. The somewhat weaker sources with complex morphology
prefer richer environments than classical doubles. 
Such radio sources can heat the gas in their environment significantly, which is now believed to play a major role in keeping
hot atmospheres of groups and clusters of galaxies hot \citep[e.g.,][]{mghk-mc08,mghk-TS15}, and thus impacts on cooling
of hot gas, gas inflow and star-formation in massive galaxies \citep{mghk-C06}.

\begin{figure}
  \centering
  \includegraphics[width=0.7\textwidth]{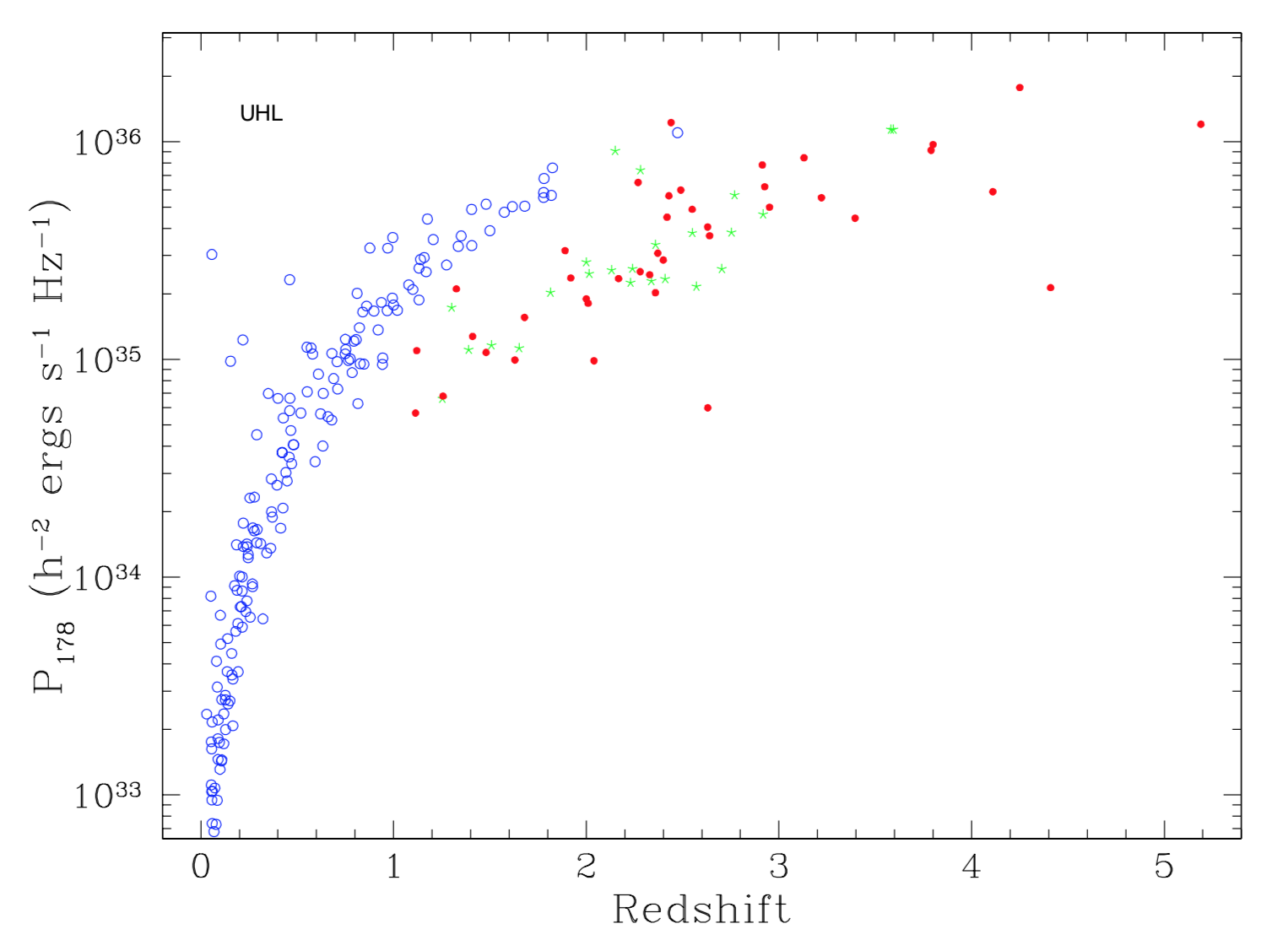}
  \caption{\small Radio power at a rest-frame frequency of 178~MHz versus redshift for the 3C sample (open circles) 
  	and fainter samples  (filled circles and stars), adopted from \citet{mghk-Carilli01}.}
   \label{mghk-f02}%
\end{figure}

By the end of the 1950s, catalogues of radio sources became available (2C, 3C, Sidney~85~MHz).
The source counts ruled out a steady-state Universe and clearly favoured big-bang cosmology \citep{mghk-norris17}. 
The important result reflected the fact that radio sources become more powerful with redshift. The most powerful radio sources
are found at cosmological redshifts \citep[e.g.,][Fig~\ref{mghk-f02}]{mghk-Carilli01}.
Because the assembly of galaxies and galaxy clusters proceeds hierarchically, bottom up, this means that radio sources must have
had an enormous
impact at high redshift. Powerful double radio sources like Cygnus~A have an energy budget of the order of $10^{62}$~erg
\citep{mghk-k05}. A reasonable fraction of this energy is transfered to the ambient gas \citep{mghk-HK13}. While this is
only a few per cent of the binding energy of the intracluster medium in the nearby host galaxy cluster of Cygnus~A,
it may be similar to or even exceed the values for a high redshift environment.  This is in line with expectations
for "pre-heating" of galaxy clusters \citep{mghk-mc08}, and may terminate star formation in the Megaparsec scale environment
\citep{mghk-rj04}.

High quality, high resolution radio maps became available with the Very Large Array (VLA) and very long baseline interferometry 
(VLBI). They showed clear detections of the jets, collimated, relativistic plasma beams that connect the energy generation
region around the supermassive black holes to the sometimes Megaparsec-scale radio lobes (Fig.~\ref{mghk-f01}). 
Such a mechanism had been proposed 
before, because of the huge energy requirements to power the lobes \citep{mghk-br74,mghk-s74}. This gave important support to 
supermassive black hole hypothesis and was a striking example that properties of large-scale radio sources can be used to constrain 
the central engine.

Here I would like to review some main points on what we have learnt from radio sources so far in the context of heating of 
intracluster gas, impact at high redshift and the central engine.

\section{Heating of the intracluster medium}\label{mghk-s:heating}
\begin{figure*}
  \centering
  \includegraphics[width=0.7\textwidth]{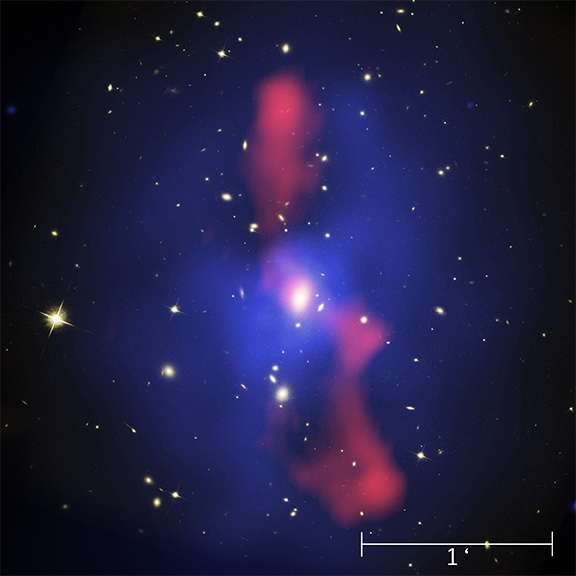}
  \caption{\small
  	Galaxy cluster MS0735.6+7421. White: optical (Hubble), blue: X-ray (Chandra), red: 330~MHz radio image.
  	The central, biggest galaxy in the cluster produces the radio source, which has displaced the X-ray gas, carving out
  	cavities. Credit: X-ray: NASA/CXC/Univ. Waterloo/B.McNamara; Optical: NASA/ESA/STScI/Univ. Waterloo/B.McNamara; 
  	Radio: NRAO/Ohio Univ./L.Birzan et al. \citep[compare also ][]{mghk-mn07}.}
   \label{mghk-f03}%
\end{figure*}

Galaxy clusters are filled with hot gas with temperatures in excess of 1~keV
\citep[for a recent review]{mghk-bw10}. A significant fraction of galaxy clusters
has cooling times in their core that are shorter than the Hubble time (cool core clusters), 
which is now thought to be offset by heating via thermal conduction and energy input by 
intermittent radio source production in the central dominant (CD) galaxy of the cluster
\citep[Fig.~\ref{mghk-f03}]{mghk-mn07,mghk-f12,mghk-TS15}.
We first review basic radio source physics and will discuss
different modes of heating from this perspective.

\subsection{Basic radio source physics}
Differential rotation in accretion flows on to a black hole will shear and amplify 
the magnetic field in the direction of rotation. This builds up magnetic pressure which can drive
an outflow along the spin axis \citep[for a recent review]{mghk-Pea12}. The toroidal magnetic field will at the same time
accelerate and collimate the outflow to a finite opening angle. 
Simulations have produced half opening angles of $3-7^\circ$ for the case of geometrically 
thin Keplerian accretion discs \citep[e.g.,][]{mghk-pf10}.
General relativistic simulations of jet formation from geometrically thick accretion tori show 
somewhat wider jets. For example, \citet{mghk-b09} find about $20^\circ$.
The magnetic field drops strongly outwards, away from the jet formation region while
the kinetic energy increases \citep{mghk-Pea12}, dominating after a few tens of Schwarzschild radii
\citep{mghk-pf10}. 

\begin{figure*}
  \centering
  \includegraphics[width=0.8\textwidth]{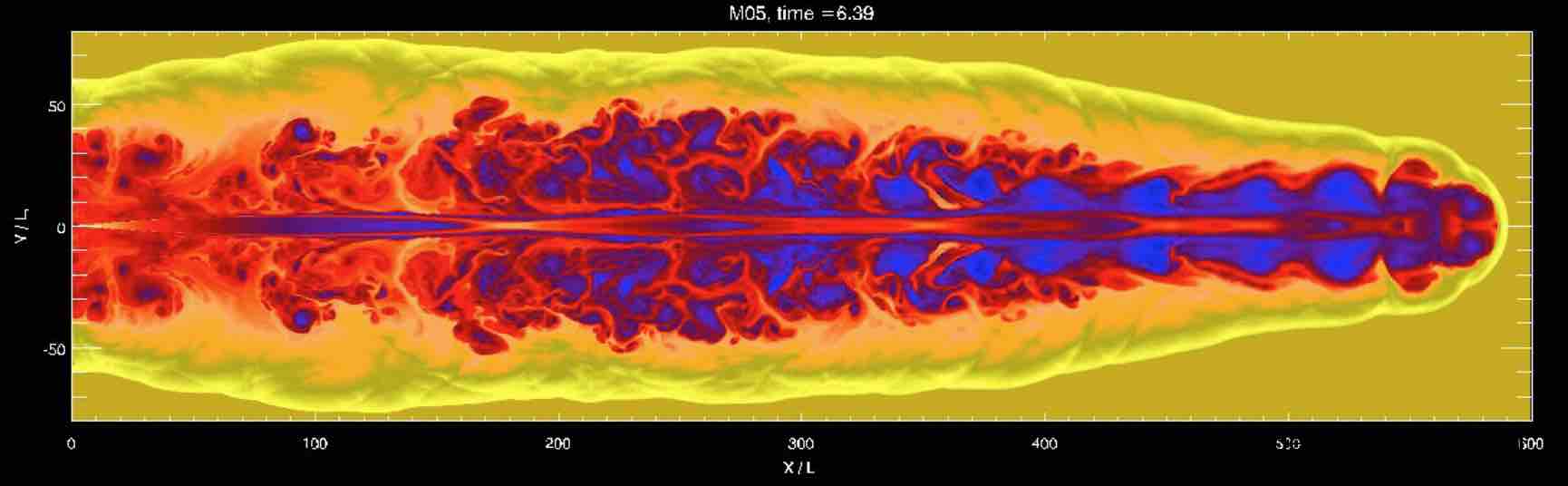}
  \includegraphics[width=0.8\textwidth]{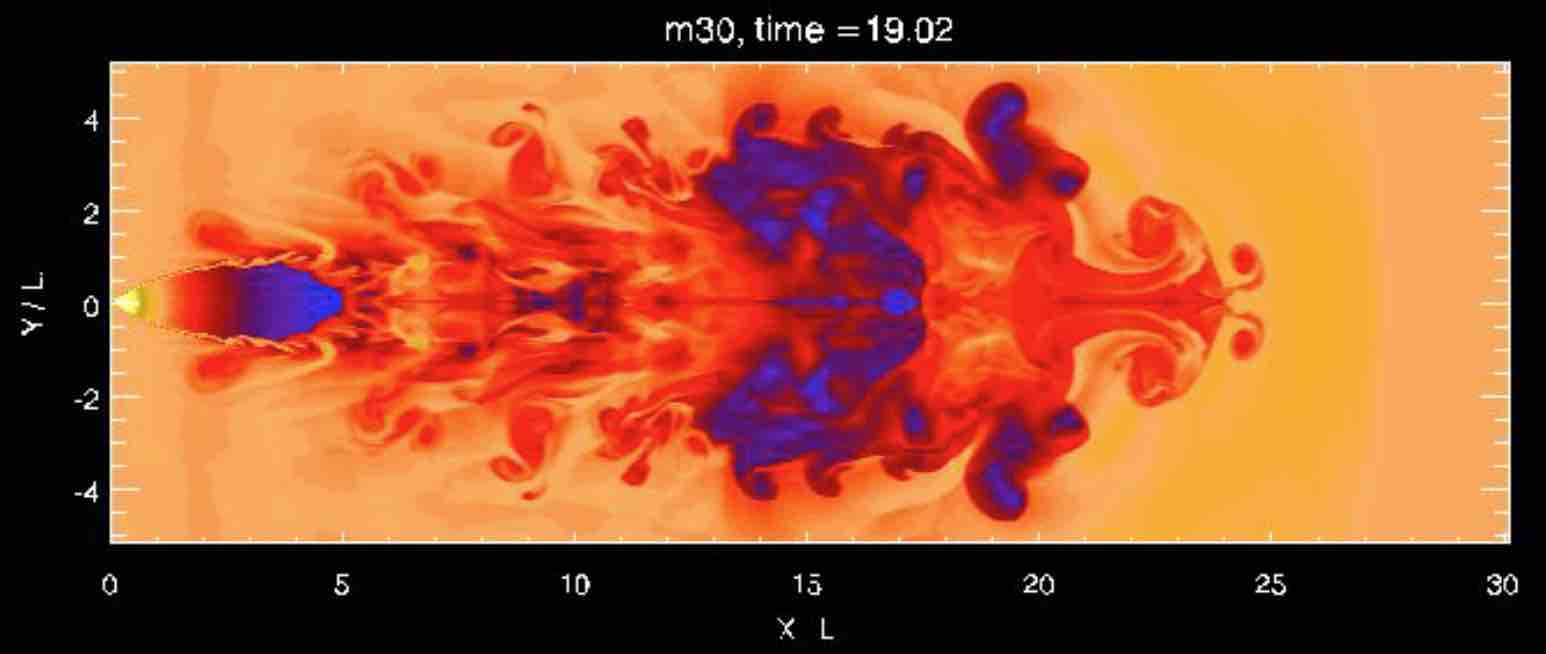}
  \caption{\small
  	Hydrodynamic simulations of jets with different opening angles. Top: With a half-opening angle of 
  	$5^\circ$, jets may collimate hydrodynamically. They stay supersonic up to the tip of the cocoon,
  	and may drive their terminal hotspots to infinity as long as the black hole remains active. 
  	Bottom: Jet with a half-opening angle of 30$^\circ$. The jet cannot collimate. Instead the terminal hotspot
  	stalls where the jet's ram pressure equals the ambient pressure. Downstream, a transsonic, strongly entraining
  	flow evolves.}
   \label{mghk-f04}%
\end{figure*}

Because of the still finite opening angle, the ram pressure in the jet now drops as $r^{-2}$, $r$ being the
distance to the black hole. The jet can recollimate hydrodynamically to zero opening angle,
when the sideways ram pressure has dropped to the level of the ambient pressure. 
\citet{mghk-kea12} show for a constant ambient gas distribution
that this depends solely on the half-opening angle of the initially conical jet (see also Fig~\ref{mghk-f04}):

If it is less than $24^\circ$, the jet will collimate hydrodynamically at a length scale $L_\mathrm{1a}$,
which, for a jet at half the speed of light, a half-opening angle of $5^\circ$, jet power $Q_0$ and
ambient pressure $p_\mathrm{x}$ is given by: 
\begin{equation}
L_\mathrm{1a} = 3 \,\mathrm{kpc} \left(\frac{Q_0}{10^{45}\mathrm{\,erg\,s^{-1}}}\right)^{1/2}
	\left(\frac{p_\mathrm{x}}{10^{-9}\mathrm{\,dyn\,cm^{-3}}}\right)^{-1/2}\, .
\end{equation}
The jet then also becomes 
underdense with respect to its environment and forms a cocoon of shocked jet material around the
jet, which will often provide the pressure for the hydrodynamic recollimation. 
The jet lobe system then expands self-similarly, length scales expanding as ($t$: time, 
ambient density: $\rho_0 (r/r_0)^\kappa$):
\begin{equation}
L\propto \left(\frac{Q_0 t^3}{\rho_\mathrm{0}}\right)^\frac{1}{\kappa+5}
\end{equation}
 \citep{mghk-ka97,mghk-k03}, until the lobes reach pressure 
equilibrium with the environment \citep{mghk-HK13}.  The source can however continue expanding as long as the jet is active.

If the half-opening angle is greater than $24^\circ$, then the jet cannot recollimate before the hotspot stalls
because the forward ram pressure is matched by the ambient pressure. Such sources will have a strong shock
near
\begin{equation}
L_\mathrm{1c} = 2 \,\mathrm{kpc} \left(\frac{Q_0}{10^{43}\mathrm{\,erg\,s^{-1}}}\right)^{1/2}
	\left(\frac{p_\mathrm{x}}{10^{-11}\mathrm{\,dyn\,cm^{-3}}}\right)^{-1/2}\, ,
\end{equation}
where we have now assumed a half-opening angle of $30^\circ$. The flow will then evolve transsonically,
entraining significant amounts of ambient material.

\citet{mghk-kea12} have proposed that these two basic radio source morphologies may explain the observed
Fanaroff-Riley (FR) dichotomy. Jets with wide opening angle would produce the centre-bright FR~I radio sources, whereas
jets with narrow opening angles would form the edge-brightened FR~IIs. Other FR~I models have been proposed, based
on instabilities \citep[e.g.,][]{mghk-kb07}, expansion of overpressured jets \citep{mghk-pm07}, and entrainment
of stellar winds \citep{mghk-k94}.

\begin{figure*}
  \centering
  \includegraphics[width=0.9\textwidth]{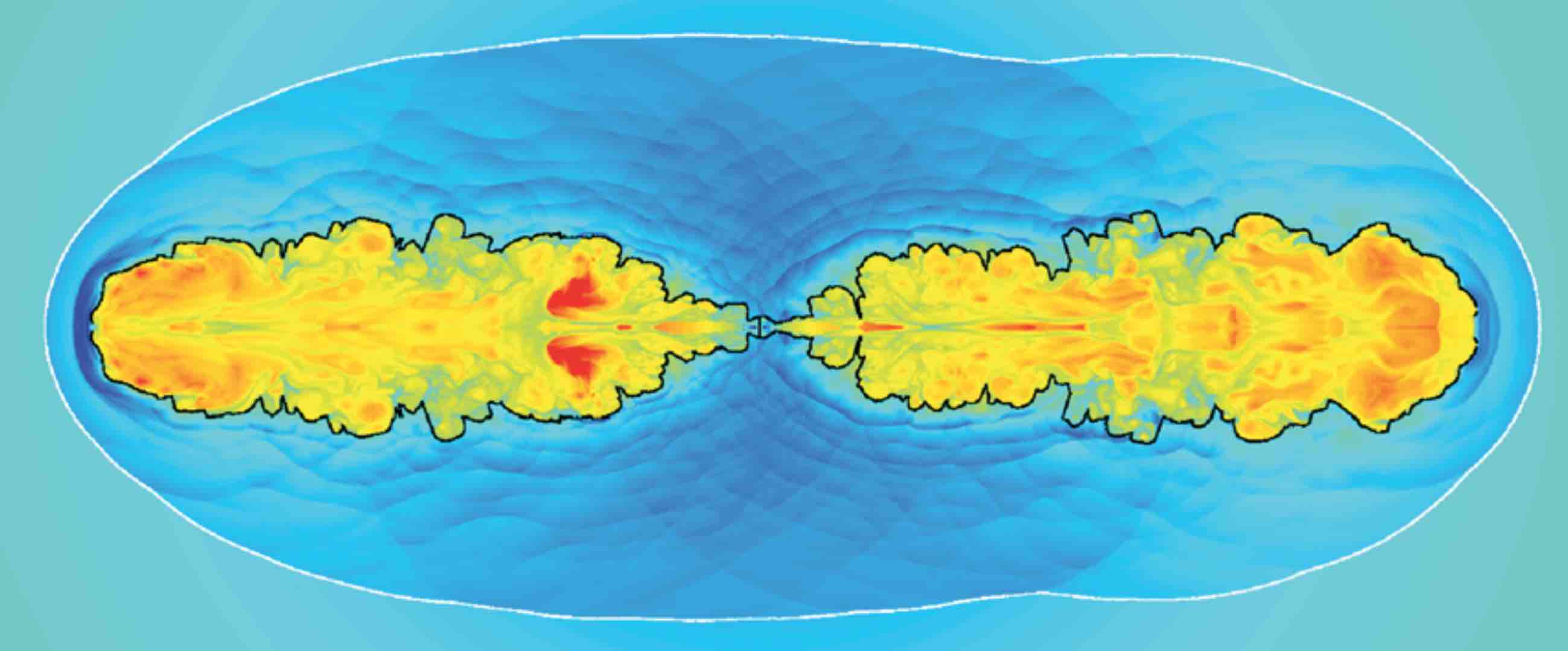}
  \caption{\small 
  Simulation of the interaction of a radio lobe with dense intracluster medium, adopted from \citet{mghk-HK13}.
  The outer bow shock is seen as a white line, the contact surface that separates the shocked ambient gas from
  the shocked jet gas in the cocoon / radio lobe is shown as a black line. The lobes have started to detach from the 
  central region, allowing ambient gas to refill the cavity. Ripples are seen in the shocked ambient gas
  due to motion of the contact surface due to the turbulent motions in the cocoon / radio lobe region.}
   \label{mghk-f05}%
\end{figure*}
When the pressure of the lobes approaches the ambient pressure, the lobes start to detach from each other and the
ambient gas refills the centre of the cavity (Fig.~\ref{mghk-f05}).

\subsection{Modes of heating}
Irreversible heat transfer is conveniently measured by the entropy index $s=kT n^{-2/3}$, which is constant under adiabatic
changes, but increases in shocks and by viscous dissipation.

\subsubsection{Strong shocks}
The most obvious heat source for intracluster gas is the leading bow shock of a radio source. 
Strong shocks are usually not seen in observed sources, as expected from radio source models: 
Since the Mach number of the bow shock quickly declines, the observation only implies
that radio sources tend to live much longer than they need to get to pressure equilibrium. Only the innermost regions of a
cluster will then be efficiently heated by the shock, and much of the energy is not used for heating the cluster gas.
The simulations of \citet{mghk-HK13} illustrate this process (Fig.~\ref{mghk-f05}). The simulated radio source
heats the central 50~kpc very efficiently. Because of this, that gas is buoyantly uplifted and the region is refilled with
gas with only 10-20~per cent higher entropy index. To prevent significant cooling in this way, the radio source
must be active recurrently. Observations find the rising bubbles left-over in the X-ray gas \citep[e.g.,][]{mghk-f12}.
So this may indeed be a mode that is frequently realised in nature.

\subsubsection{Weak shocks / ripples}
A similar faction of the energy in a radio source is stored in lobes and, respectively, ambient gas
\citep[e.g.][]{mghk-HK13,mghk-EHK16}. About half of the lobe energy is accounted for by turbulent motions.
The random motions disturb the contact surface between radio lobes and intracluster medium and run weak shock waves
or ripples into the part of the intracluster medium that has already been shocked by the leading bow shock (Fig.~\ref{mghk-f05}).
Such ripples are observed in some galaxy clusters and may carry enough energy to offset cooling in the cluster
\citep{mghk-f12}. The observed sources suggest that the perturbations of the contact surfaces are enhanced by
intermittency of the radio source or, perhaps, precession (more below). Viscous dissipation of such ripples could heat
the intracluster gas to larger radii than the strong shocks.

\begin{figure*}
  \centering
  \includegraphics[width=0.9\textwidth]{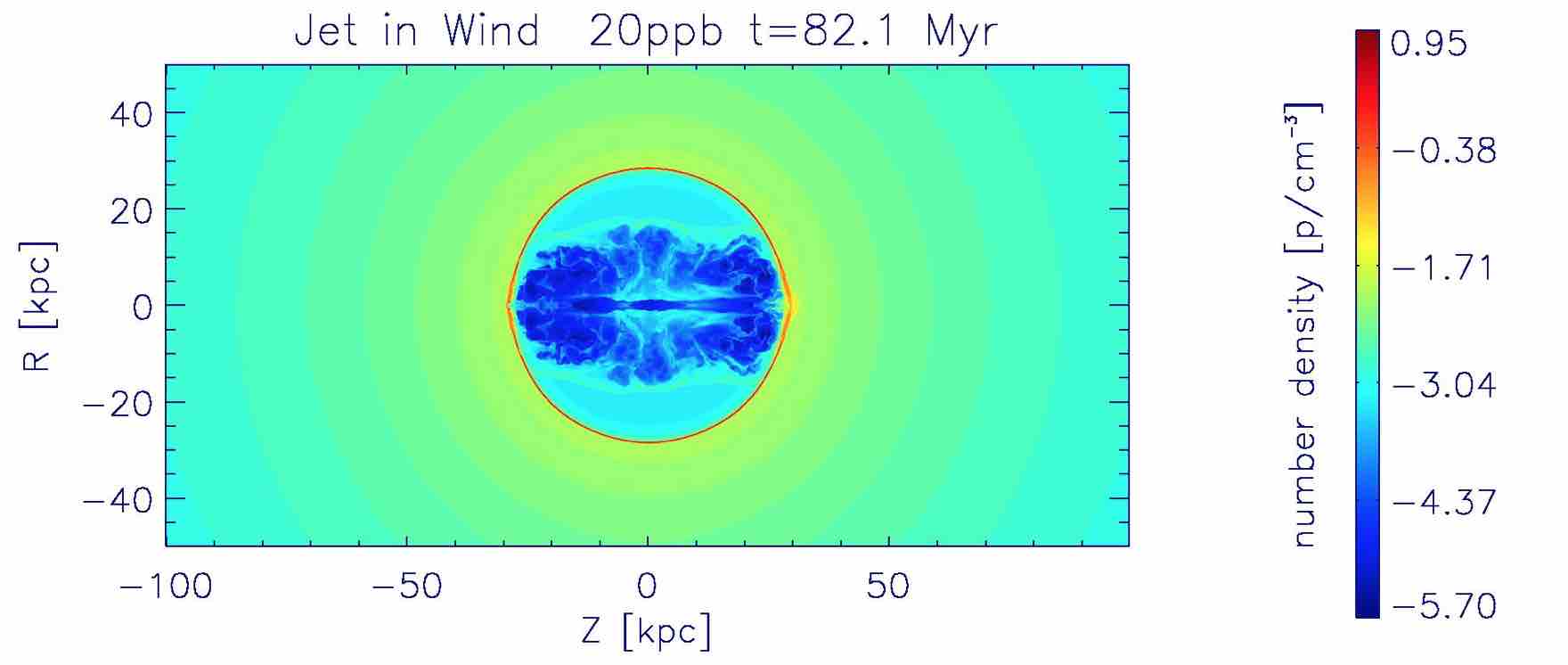}
  \includegraphics[width=0.9\textwidth]{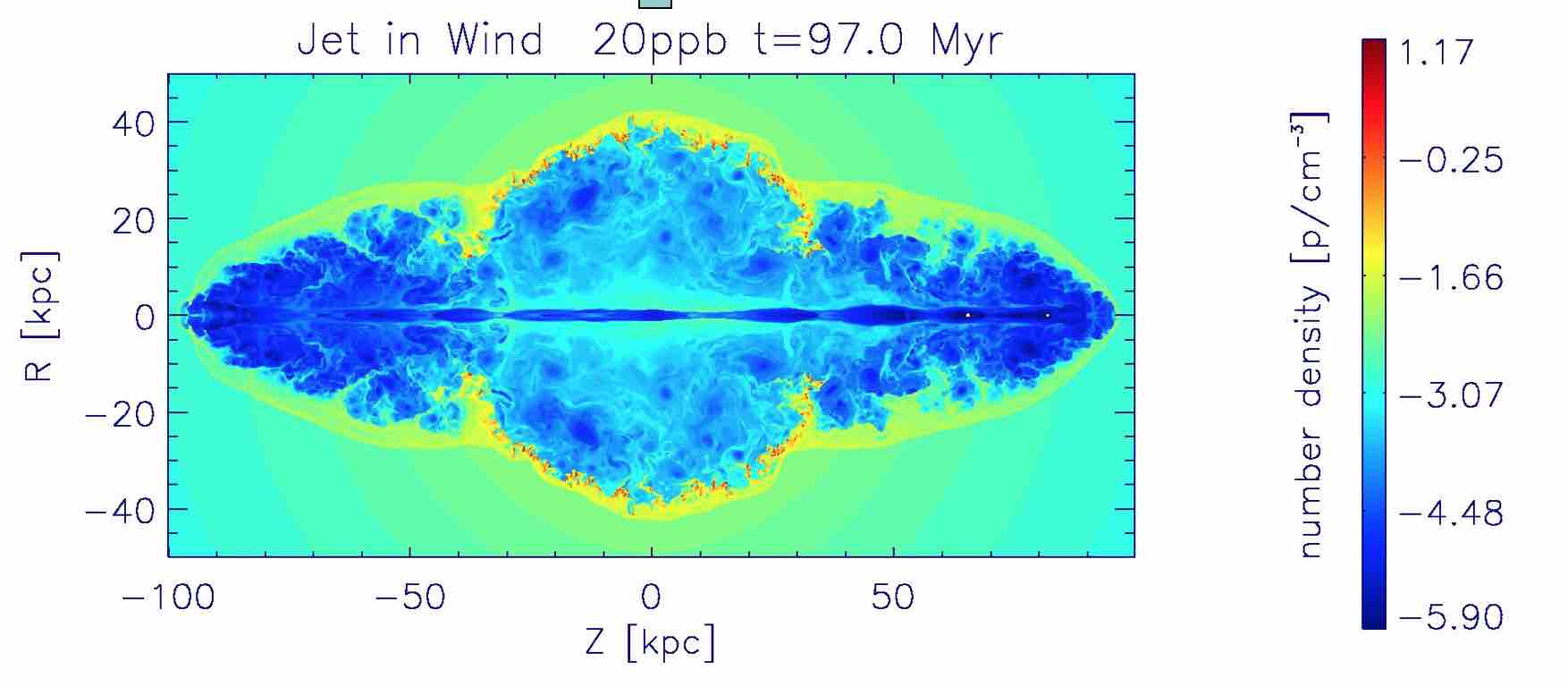}
  \caption{\small
  	Hydrodynamic simulations of a jet in a galactic wind shell \citep{mghk-k05b}. The jet is efficiently trapped by the wind shell
  	until the radio lobe fills the shell region. The shell is then accelerated and fragmented by the Rayleigh-Taylor instability.
  	This may explain Lyman~$\alpha$ self-absorption in high redshift radio galaxies.}
   \label{mghk-f06}%
\end{figure*}

\section{High redshift radio sources}
High redshift radio sources (HZRGs, redshift $z>1$) can be more powerful than their low redshift counterparts and are associated
with larger emission line lobes and very high rates of star formation \citep[for a review]{mghk-md08}.
Most HZRGs live in so-called proto-clusters, regions of several times $10^{14} M_\odot$
that are not virialised yet, but are thought to evolve into big clusters of galaxies by redshift zero \citep{mghk-Vea07}.
The space density of galaxy clusters at redshift zero is compatible with the idea that all clusters had such a powerful
radio source when they where assembled \citep{mghk-mld01}. HZRGs hence highlight the early stages of galaxy clusters
and at the same time are laboratories for feedback in massive galaxies when they have not yet transformed into 
red ellipticals with little star formation.

\subsection{Lyman $\alpha$ haloes}
HZRG are frequently associated with emission line haloes with sizes up to 100~kpc \citep{mghk-md08}.
High fidelity radio images of HZRGs are not yet available, but the halos are aligned in direction 
with the radio sources and may thus correspond to the region occupied by the radio lobes.
They evidence massive gaseous outflows with turbulent and bulk outflow velocities of the order of a few hundred
km/s \citep{mghk-Nea17}. While energetically feasible, it remains unclear if the radio source is solely responsible
for these outflows. In hydrodynamic simulations of the interaction of jets with clumpy, dense interstellar medium in the 
host galaxies, \citet{mghk-Mea16} find efficient energy transfer, but only for as long as the jet is contained in the galaxy,
which is too short to produce the large-scale outflows. In multi-phase turbulence simulations, \citet{mghk-ka07}
show that the turbulence of the radio plasma in jet cocoons could accelerate entrained emission line gas to the
observed velocities. It is well possible that early jet and also stellar feedback play a crucial role in launching the gas off the galaxy
with cocoon turbulence taking over on larger scales. 

\subsection{Lyman $\alpha$ self-absorption}
Lyman $\alpha$ emission from HZRG is frequently self-absorbed within typically a few hundred km/s bluewards of the 
emission peak \citep{mghk-md08,mghk-Swea15}. The best explanation is an expanding, radiative shell in the snow-plough phase
\citep{mghk-k05b}. The shell cannot be driven by the radio source, because it would then be too fast, probably would not cool enough
to form neutral hydrogen, and, if it would, its Ly~$\alpha$ luminosity would violate observational limits.
The shell is therefore likely driven by the stellar feedback in the strongly star-forming hosts. It sweeps up the hot atmosphere
of its host galaxy, likely transforming it to stars. This might be a significant channel of forming globular clusters
directly in the halo of a galaxy \citep{mghk-k02}. The wind shells are disrupted by the radio sources 
\citep[Fig~\ref{mghk-f06}]{mghk-k05b}

\begin{figure*}
  \centering
  \includegraphics[width=0.3\textwidth]{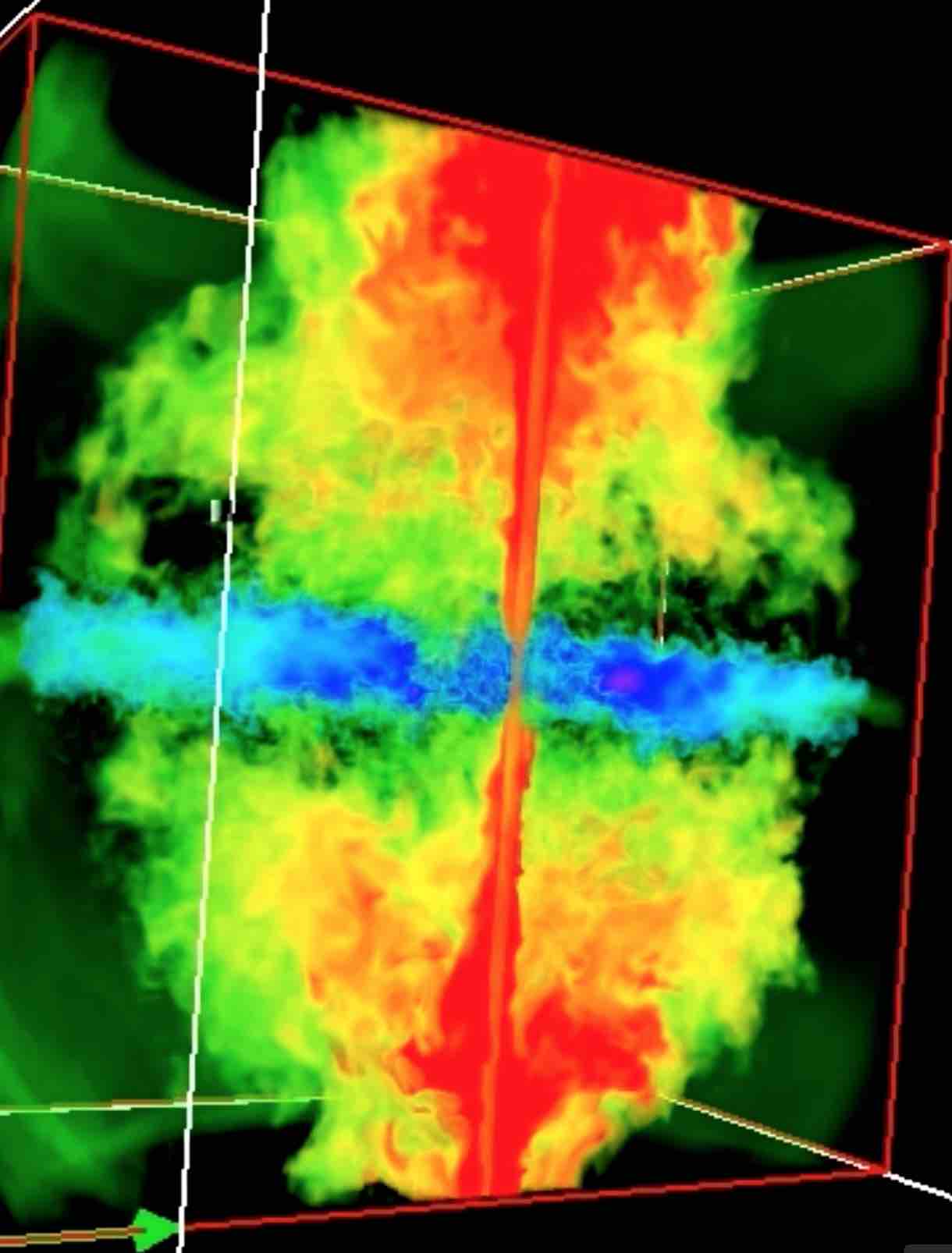}
  \includegraphics[width=0.6\textwidth]{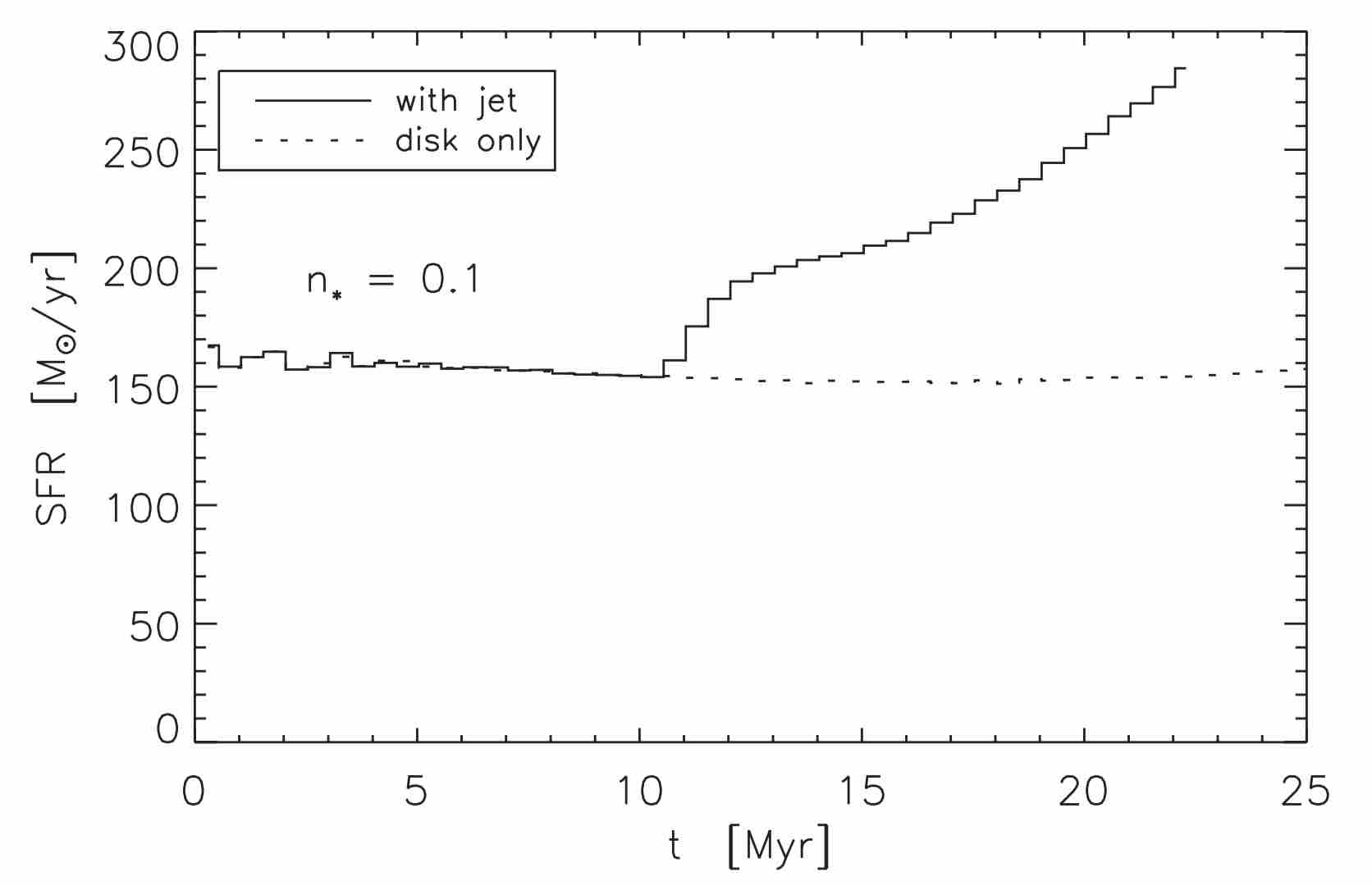}
  \caption{\small
  	Hydrodynamic simulations of jet induced star formation in a massive high redshift galaxy \citep{mghk-Gea12}. The radio lobes
  	surround the interstellar medium and squash it by their high pressure. This enhances the star formation rate significantly.}
   \label{mghk-f07}%
\end{figure*}
\subsection{Jet-induced star formation and galaxy transformation}
Powerful HZRGs provide excess pressure over a period of at least tens of Myr. This may further compress any dense clouds affected 
by the radio lobes, in particular in the host galaxy and enhance the star formation rate. We have tested this quantitatively in 
hydrodynamic simulations \citep[Fig~\ref{mghk-f07}]{mghk-Gea12}, finding about a factor of two increase. Subsequent instabilities
may enhance the star formation rate over an even longer timescale \citep{mghk-Bea16}. This raises the question,
if radio source activated, boosted star formation is preferentially picked up in observations at high redshift \citep{mghk-SM12}.
 
The radio source thus transforms the host galaxy in three ways: Firstly, the gas in the galaxy is used up much faster than without
the radio source. Secondly, any dense gas around the galaxy (e.g., cold streams or wind shells) will suffer the same fate.
Thirdly, the hot gas that has not been swept up previously by star-formation driven winds will now be heated to the multi-keV
regime, increasing its cooling time by orders of magnitude. Such a galaxy must, no doubt, quickly become a red and dead 
elliptical.

\section{The central engines}

\subsection{Rotational energy extraction and jet composition}
Jets carry properties of the central engine out to large scales, where they may be observed with comparative ease.
Energy, momentum and mass flux, plasma composition and the large-scale electric current 
 through the jet-lobe system \citep{mghk-gkc09}
are conserved quantities which characterise the processes in the direct vicinity of supermassive black holes.
For example, modelling the jet power in MS0735.6+7421 (Fig.~\ref{mghk-f03}) and comparing with the likely accretion power
of the supermassive black hole in the host galaxy, \citet{mghk-McN09} proposed that the radio source
may be powered by extraction of rotational energy of the supermassive black hole \citep{mghk-bz77}.
This is one of the motivations to model the kinetic power. For a given jet power, the observed radio luminosity
depends on the environment \citep[e.g.,][]{mghk-HK13,mghk-HK14,mghk-EHK16}.
Radio fluxes, sizes and spectra, in combination with X-ray data for the environment or, in the absence of the latter,
optical magnitudes of the host galaxies as a proxy, allow to model composition and kinetic power of individual
radio sources in detail. We have recently applied such techniques to a sample of powerful FR~II radio sources
\citep{mghk-t18a,mghk-t18b}. The distribution of jet powers extends up to about $10^{47}$~erg/s, the Eddington luminosity
of a $10^9 M_\odot$ black hole. Hence, there is currently  no evidence of spin powering for this population.
Further, our results are consistent with pure electron-positron plasma in the radio lobes. In this context, it is interesting to note
  \begin{figure}
  \centering
  \includegraphics[width=0.8\textwidth]{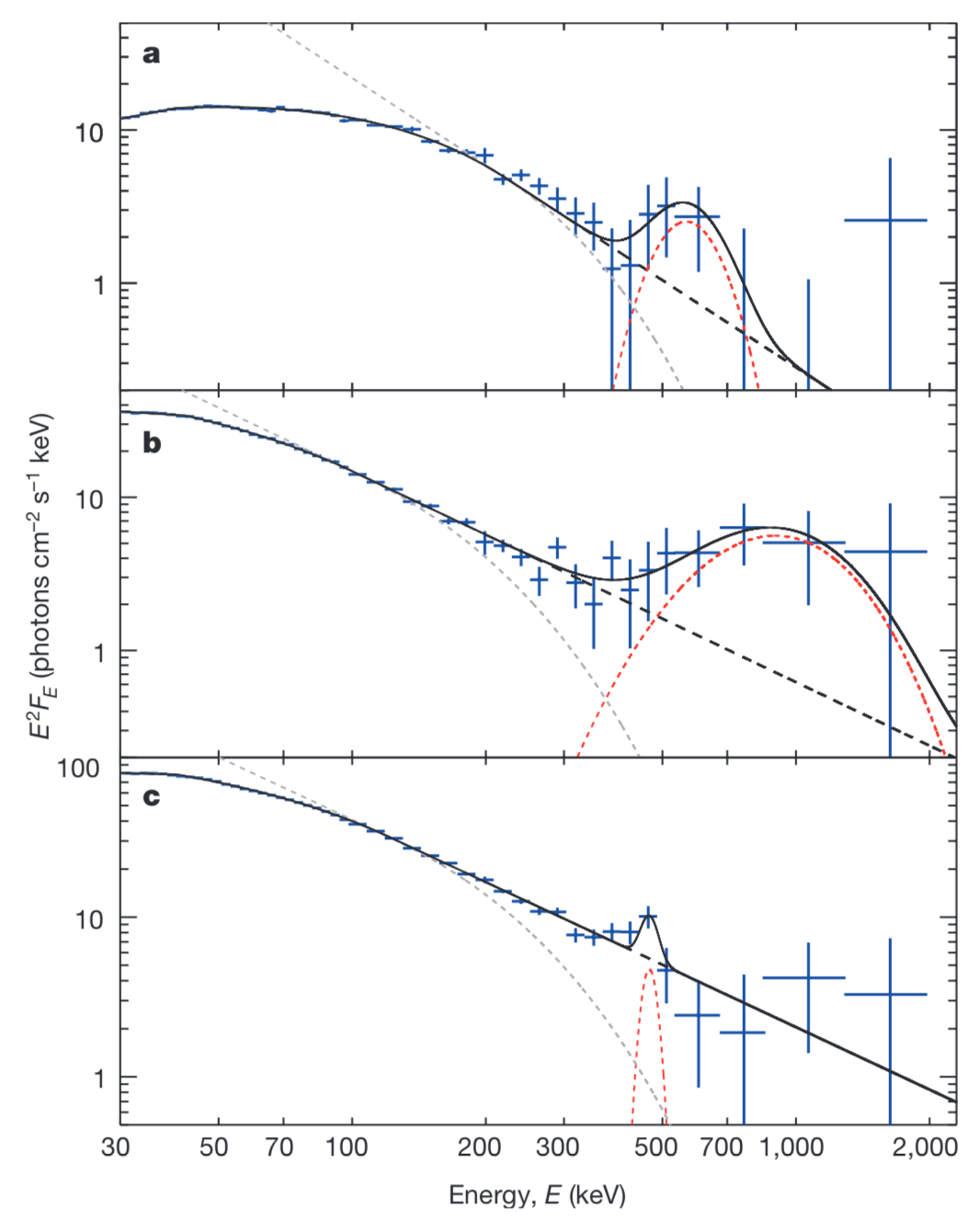}
 \caption{\small
  	Evidence for electron-positron plasma from the galactic, jet-producing stellar mass black hole in V404~Cygni \citep{mghk-s16a}. 
  	A broad annihilation line with variable width has been detected with the gamma ray spectrometer onboard INTEGRAL.}
    \label{mghk-f08}%
\end{figure}
that the 511~keV electron-positron annihilation feature has recently
been directly observed in a jet production event of a galactic, stellar mass black hole
\citep[Fig.~\ref{mghk-f08}]{mghk-s16a}. Past jet production of the Milky Way's supermassive black hole, Sgr~A*, of which the Fermi bubbles
may be evidence of, may have contributed to the Galactic positron reservoir \citep{mghk-s16b}. 

\subsection{Binary supermassive black holes}
Extragalactic radio sources are ideally suited to track binarity in supermassive black holes \citep{mghk-bbr80}.
After a galaxy merger, the two supermassive black holes are expected to quickly approach each other to sub-parsec separation
\citep{mghk-m17}. The black hole spins will then precess around the orbital angular momentum vector with the geodetic period:
 \begin{equation}
 P_\mathrm{gp} = 41\,\mathrm{Myr}\, d_\mathrm{pc}^{5/2} M_9^{-3/2} r^{-1} (4-r)^{-1}\, ,
 \end{equation}
where $d_\mathrm{pc}$ is the separation in pc, $M_9$ is the total black hole mass in $10^9 M_\odot$, and  $r$ 
 is the mass ratio of the two black holes. The timescale is comparable to the age of extended radio sources, and hence, if the jet is emitted along the spin axis of one black hole, it's precession might be observable. Associated orbital timescales of the order of tens or hundreds 
 of years are accessible to VLBI multi-epoch observations.

There is indeed plenty of evidence for precession in radio sources (Krause et al. in prep.). 
The perhaps best studied example, Cygnus~A
shows point-reflection symmetry, the jet is not identical with the lobe axis, and the direction of jet and counterjet
differ by 178$^\circ$, difficult to explain in a standard jet plus counterjet scenario or by jet-cloud interactions, 
but perhaps expected from relativistic aberration in precessing jets \citep{mghk-g82}. The geometry is consistent with a precession period of $\approx 10^6$~yr, which can be explained by a binary supermassive black hole with less than 0.5~pc separation
(Krause et al. in prep.).
The VLBI jet has a helical structure \citep{mghk-Boea16a} which can be explained by orbital motion of a binary with about
0.05~pc separation, consistent with the large-scale analysis. 

We analysed a complete sample of radio sources (Krause et al. in prep.). Point symmetry, relativistic aberration
ring-like hotspots and other features relating to precession are frequently seen.
We find good evidence for precession in 73~per cent of the 33 sources.

Complementary evidence for binary black holes in jet sources is also very strong. For example,
\citet{mghk-l13} show that VLBI jets commonly change their position angle in time, some show periodic changes,
as expected from orbital motion in a jet-producing binary.

These findings strongly suggest that powerful extragalactic radio sources may point to close binary supermassive black holes,
with implications for the cosmological history of black hole merging and gravitational wave experiments.
\begin{figure}
  \centering
  \includegraphics[width=0.9\textwidth]{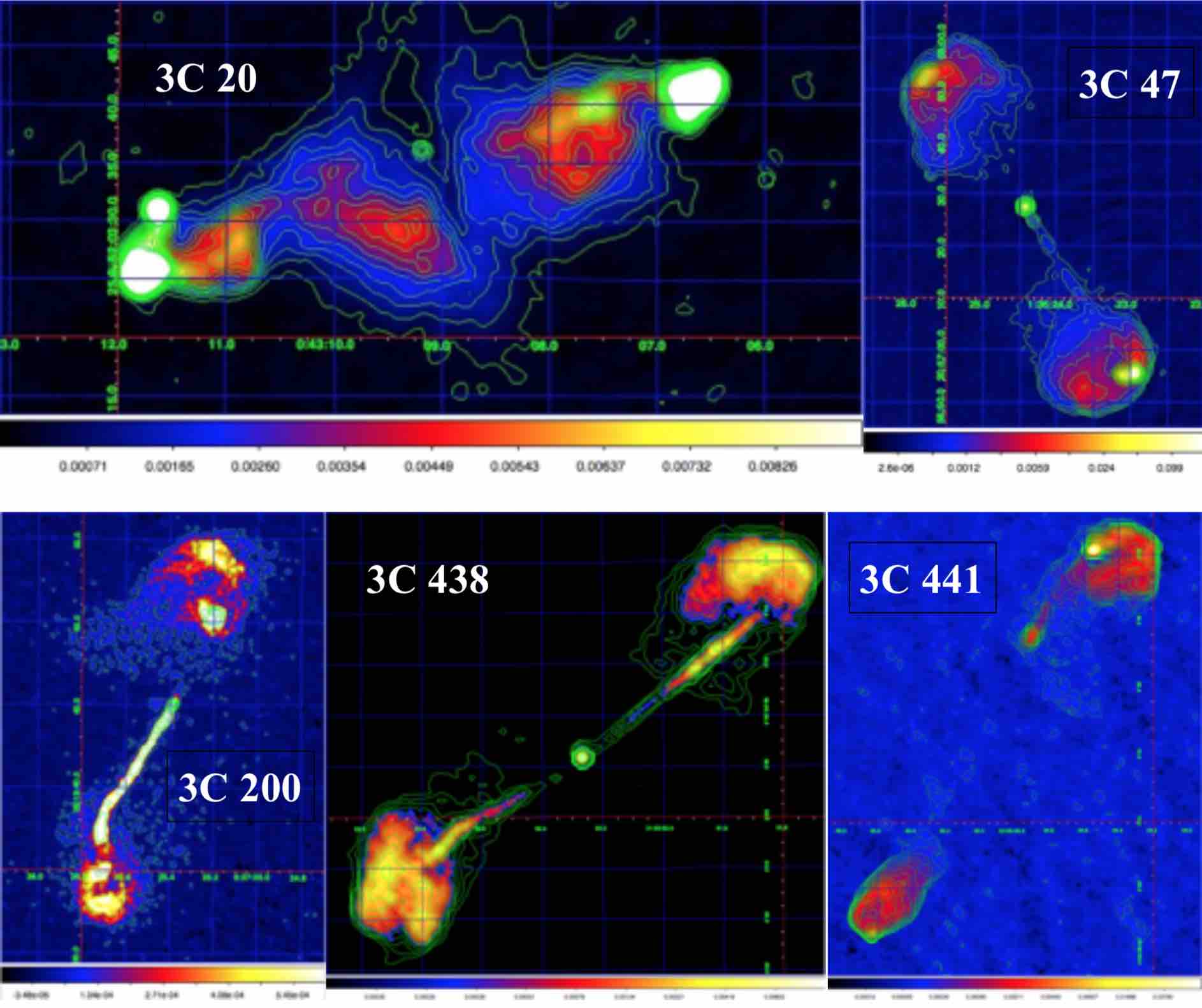}
  \caption{\small
  	VLA radio maps of radio sources with strong evidence for jet precession.}
   \label{mghk-f09}%
\end{figure}

%

\end{document}